\documentstyle[12pt]{article}
\newlength{\extraspace}
\newlength{\extraspaces}
\newcommand{\be}{\begin{equation}
\addtolength{\abovedisplayskip}{\extraspaces}
\addtolength{\belowdisplayskip}{\extraspaces}
\addtolength{\abovedisplayshortskip}{\extraspace}
\addtolength{\belowdisplayshortskip}{\extraspace}}
\newcommand{\ee}{\end{equation}}
\newcommand{\ba}{\begin{eqnarray}
\addtolength{\abovedisplayskip}{\extraspaces}
\addtolength{\belowdisplayskip}{\extraspaces}
\addtolength{\abovedisplayshortskip}{\extraspace}
\addtolength{\belowdisplayshortskip}{\extraspace}}
\newcommand{\ea}{\end{eqnarray}}
\begin{document}
\thispagestyle{empty}
\begin{flushright}
SIT--LP--02/09 \\
{\tt hep-th/0209165} \\
September, 2002
\end{flushright}
\vspace{7mm}
\begin{center}
{\large \bf Superon-graviton model and supersymmetric structure \\
            of spacetime and matter 
\footnote{
Invited Talk at Advanced Study Institute, Praha-Spin-02, 
Praha, Czech Republic, July 14-27, 2002} \
} \\[20mm]
{\sc Kazunari Shima}${}^{\rm a}$\footnote{
\tt e-mail: shima@sit.ac.jp}, \
{\sc Motomu Tsuda}${}^{\rm a}$\footnote{
\tt e-mail: tsuda@sit.ac.jp} \
and \
{\sc Manabu Sawaguchi}${}^{\rm b}$\footnote{
\tt e-mail: sawa@sit.ac.jp}
\\[5mm]
${}^{\rm a}${\it Laboratory of Physics, 
Saitama Institute of Technology \\
Okabe-machi, Saitama 369-0293, Japan} \\[3mm]
${}^{\rm b}${\it High-Tech Research Center, 
Saitama Institute of Technology \\
Okabe-machi, Saitama 369-0293, Japan} \\[20mm]
%
\begin{abstract}
A new Einstein-Hilbert type (SGM) action describing gravitational interaction of 
Nambu-Goldstone(N-G) fermion of nonlinear supersymmetry(NL SUSY) is obtained by performing 
the Einstein gravity analogue geometrical arguments in high symmetric  four dimensional (SGM) spacetime. 
All elementary particles except graviton are regarded as the composite eigenstates of SO(10) super-Poincar\'e 
algebra(SPA) composed of the fundamental N-G fermion ``superons'' of NL SUSY. 
Some phenomenological implications of the composite picture of SGM, the linearlization of SGM and 
$N=2$ Volkov-Akulov model are discussed. 
\end{abstract}
\end{center}

\newpage


\section{Introduction}
Despite the success of the standard model(SM) as a unified model 
for the strong and the electroweak interaction, there  still remain many unsolved problems, 
e.g. it can not explain the particle quantum numbers $(Q_{e},I,Y,color, i.e. 
U(1) \times SU(2) \times SU(3) \ gauge \ structure)$, the three-generations structure and 
contains more than 28 parameters (even disregarding the mass generation mechanism for neutrino). 
The gravitational interaction is not considered. 
SM and GUT equipped minimally with supersymmetry(SUSY) 
have improved the situations in some points,  
but  they are still pathological on the proton decay problem and less predictive 
due to more than 100  parameters.   \\
Although SUSY\cite{WB} is an essential notion to unify  spacetime and matter, 
unfortunately  SO(8) SUGRA in four dimensional spacetime is too small to accommodate all observed particles 
as elementary fields.  
The straightforward extension to SO($N \ge 9$) SUGRA has a difficulty 
due to so called the no-go theorem on the coupling of gravitation and 
the ${massless}$ ${elementary}$ high spin$(>2)$ gauge field.   \\
However it is well known that in the monopole ${phase}$, 
(i.e. at the very short distances of spacetime), 
the degrees of freedom (dimensions) of spacetime are ${fused}$ with the dimensions of 
(the linear representation of) the local symmetry, 
which allows to define a unified (composite) field strength of the monopole configuration 
through the symmetry breaking $SU(2) \times SO(3,1) 
\rightarrow U(1) \times SO(3,1)$\cite{PtH}.     
These phenomena  suggest that spacetime itself would reveal unfamiliar features 
at the short distance by the identification(${fusion}$) of 
the symmetries of spacetime with those of matter and that these ultimate spacetime 
would  be specified by a certain unified (composite) field  strength(curvature), 
where the  no-go theorem becomes irrelevant in a sense that the fundamental  Lagrangian 
with $N \ge 9$ SUSY may be written down. 
Also, we think that from the viewpoint of simplicity and beauty of nature 
it is interesting to attempt the accommodation of all observed  particles in ${a \ single}$ 
irreducible representation of a certain algebra(group), especially for spacetime having 
a certain boundary (boundary condition).  
The fundamental theory should be given by only the geometrical arguments of  high symmetrical spacetime 
and its spontaneous breakdown, which is encoded in the geometrical argument of spacetime by itself. 
In this talk we would like to present a model along this scenario.  \\
\section{Superon-Graviton Model(SGM)}
Among single irreducible representations of all SO(N) extended super-Poincar\'e(SP) symmetries, 
the massless irreducible representations of SO(10) SP algebra(SPA) is the only one that accommodates 
minimally all observed particles including the graviton\cite{KS1}\cite{KS2}. 
By considering that (i)for the  massless  case the  algebra  of the supercharges 
of SO(10) SPA in  the light-cone frame can be recasted as those of the 
creation and annihilation operators of fermions and (ii)10 generators  $Q^{N}(N=1,2,..,10)$ of SO(10) SPA 
are 
the fundamental representations of SO(10) internal symmetry  and 
decomposed  
$\underline{10} =  \underline 5+ \underline 5^{*}$  with respect to SU(5) following 
$SO(10) \supset SU(5) $ and span $2\cdot2^{10}$ dimensional massless irreducible representation 
of SO(10) SPA, 
we can regard 10 generators
$\underline{10} =  \underline 5+ \underline 5^{*}$ as the fundamental massless objects; 
$\it{a}$ ${superon}$-${quintet}$ and $\it{an}$ ${antisuperon}$-${quintet}$ with ${spin \ {1 \over 2}}$ 
and that all the helicity states are the massless (gravitational) ${eigenstates}$ of spacetime and matter 
with SO(10) SP symmetric structure, which are composed of $\it{superon}$. 
We regard  (broken) SO(10) SP symmetry is to spacetime and matter(nature) 
what O(4) symmetry is to the relativistic hydrogen atom. 
To survey the physical implications of ${superon}$-${graviton}$ ${model(SGM)}$ for spacetime and ${matter}$ 
we assign the following SM quantum numbers to superons and adopt the following symbols(and the conjugates 
for anti-superons). 
%
%
%
%
%
%
%
\begin{eqnarray}\underline{5}  =  \Bigr[ Q_{a}(a=1,2,3) ,Q_{m}(m=4,5)\Bigl] 
= [(\underline 3, \underline 1;-{1 \over 3},-{1 \over 3},-{1 \over 3}), (\underline 1, \underline 2;1, 0)],
\end{eqnarray}
where we have specified (${\underline{SU(3)},\underline{SU(2)}}$; electric  charges ). 
Superon-quintet satisfy the Gell-Mann--Nishijima relation;     
${Q_{e}=I_{z} + {1 \over 2}(B-L)}$.  
Accordingly all ${2 \cdot 2^{10}}$  helicity states (up to helicity 3) are specified uniquely 
with respect to ( $SU(3), SU(2)$; electric charges ).
Here we assume boldly an ideal super Higgs-like mechanism, i.e. all unnecessary 
(for SM) higher helicity states become massive by absorbing the lower helicity states 
in $SU(3) \times SU(2) \times U(1)$ invariant way  
via 
[SO(10) SPA upon the Clifford vacuum]
$\rightarrow$ [$SU(3) \times SU(2) \times U(1)$] 
$\rightarrow$ [$SU(3) \times U(1)$]. 
We have carried out the recombinations of  the helicity states 
and found surprisingly that all the massless states necessary  for 
the SM with three generations of quarks and leptons appear in the surviving 
massless states specified by the superon contents.   \\
For three generations of leptons
[$({\nu}_{e}, e)$,  $({\nu}_{\mu}, \mu)$,  $({\nu}_{\tau}, \tau)$], we take
\begin{equation}
\Bigr[(Q_{m}{\varepsilon}_{ln}Q_{l}^{*}Q_{n}^{*}),
(Q_{m}{\varepsilon}_{ln}Q_{l}^{*}Q_{n}^{*}Q_{a}Q_{a}^{*}),
(Q_{a}Q_{a}^{*}Q_{b}Q_{b}^{*}Q_{m}^{*})\Bigl]
\end{equation}
\noindent and 
for three generations of quarks [$( u, d )$, $( c, s )$, $( t, b )$],
we have ${\em uniquely}$
\begin{equation}
\Bigr[({\varepsilon}_{abc}Q_{b}^{*}Q_{c}^{*}Q_{m}^{*}),
({\varepsilon}_{abc}Q_{b}^{*}Q_{c}^{*}Q_{l}
{\varepsilon}_{mn}Q_{m}^{*}Q_{n}^{*}),
({\varepsilon}_{abc}Q_{a}^{*}Q_{b}^{*}Q_{c}^{*}Q_{d}Q_{m}^{*})\Bigl]
\end{equation}
\noindent and their conjugates respectively.  
For $SU(2) \times U(1)$ gauge bosons [ ${{W}^{+},\ Z,\ \gamma,\ {W}^{-}}$], 
$SU(3)$ color-octet gluons [${G^{a}(a=1,2,..,8)}$], 
[$SU(2)$ Higgs Boson], [$( X,Y )$] leptoquark bosons in GUTs, 
and  [a color- and SU(2)-singlet neutral gauge boson from
${{\underline 3} \times {\underline 3^{*}}}$ (called  S  boson)] we have  
${[Q_{4}Q_{5}^{*}, {1 \over \sqrt{2}}( Q_{4}Q_{4}^{*} \pm Q_{5}Q_{5}^{*}),
Q_{5}Q_{4}^{*}]}$, 
${[Q_{1}Q_{3}^{*},Q_{2}Q_{3}^{*},..]}$
%
%
%
%
%
%
${[{\varepsilon}_{abc}Q_{a}Q_{b}Q_{c}Q_{m}]}$, 
$[Q_{a}^{*}Q_{m}]$ and  ${Q_{a}Q_{a}^{*}}$, (and  conjugates) respectively.   \\
Among predicted new particles 
one lepton-type electroweak-doublet $( \nu_{\Gamma}, \Gamma^{-} )$ with spin ${3 \over 2}$ 
with the mass of the electroweak scale $( \leq Tev)$,  
one neutral gauge boson $S$  and one doubly charged lepton $E^{2-}$ 
are color singlets and can be observed directly. 
Further studies are needed to estimate the masses of  $S$ and  $E^{2-}$.   \\
More to see the potential of SGM and to survey the evidence of the compositeness of matter 
in the low energy we interpret(reproduce) the Feynman diagrams of SM(GUT) in terms of the superon pictures, 
i.e. a single line of a propagating particle is replaced by multiple lines representing superons 
in the particle under two assumptions at the vertex; 
(i) the analogue of the OZI-rule of the quark model and 
(ii) the superon number conservation. 
Many remarkable new insights are obtained qualitatively, e.g. in SM; 
naturalness of the mixing of $K^{0}$-$\overline{K^{0}}$, 
$D^{0}$-$\overline{D^{0}}$ and  $B^{0}$-$\overline{B^{0}}$, 
no CKM-like mixings among the lepton generations, 
${\nu_{e} \leftrightarrow \nu_{\mu} \leftrightarrow \nu_{\tau}}$ transitions beyond SM,  
strong CP-violation, small Yukawa couplings and no  $\mu \longrightarrow e + \gamma$ 
despite the compositeness, etc. and in (SUSY)GUT; proton is stable without R-parity by hand
(i.e.,absence of the dangerous diagrams), etc.\cite{KS1}\cite{KS2}. 
SGM may be the most economic model.  The arguments are group theoretical so far.  \\
\section{Fundamental Theory of Superon-Graviton 
         Model(SGM)} 
By noting the supercharges $Q$ of Volkov-Akulov(V-A) model\cite{VA}  of the NL SUSY  given 
by the supercurrents
${J^{\mu}(x)={1 \over i}\sigma^{\mu}\psi(x)
-\kappa \{ \mbox{the higher order terms of $\kappa$, $\psi(x)$}\}}$ 
satisfy the SP algebra,  
we find that the fundamental theory of SGM for spacetime and matter
at(above) the Planck scale is SO(10) NL SUSY in the curved spacetime.
This is the field-current identity and  justifies our bold assumption
that the generator(supercharge) $Q^{N}$ ($N=1,2,..10$) of SO(10) SPA 
in the light-cone frame represents the fundamental massless  particle, 
$superon$ $with$ $spin$ ${1 \over 2}$.
We have written down the SGM action by performing  the similar arguments to 
Einstein general relativity theory(EGRT) in  high symmetric four   dimensional (curved)  ${SGM \ spacetime}$, 
where NL SUSY N-G fermion degrees of freedom  $\psi(x)$ (i.e. the coset space coordinates of 
superGL(4R)/GL(4R) representing N-G fermions) are embedded 
at every curved spacetime point\cite{KS2}: 
\begin{equation}
L_{SGM}=-{c^{3} \over 16{\pi}G}\vert w \vert(\Omega + \Lambda ), 
\label{SGM}
\end{equation}
where 
$\vert w \vert={\rm det}{w^{a}}_{\mu}={\rm det}({e^{a}}_{\mu}+ {t^{a}}_{\mu})$, 
${t^{a}}_{\mu}=(\kappa/2i)\sum_{j=1}^{10}(\bar{\psi}^{j}\gamma^{a}
\partial_{\mu}{\psi}^{j}
- \partial_{\mu}{\bar{\psi}^{j}}\gamma^{a}{\psi}^{j})$, and 
${\kappa^{-1}={c^{3}\Lambda \over 16{\pi}G}} $ is a fundamental volume of 
four dimensional spacetime 
and $\Lambda$ is a  ${small}$ cosmological constant related to the superon-vacuum coupling constant. 
$\Omega$ is a new scalar curvature analogous to the Ricci scalar curvature $R$ of EGRT, 
whose explicit expression is obtained  by just replacing ${e^{a}}_{\mu}(x)$  
by ${w^{a}}_{\mu}(x)$ in Ricci scalar $R$. 
These results can be understood intuitively by observing that 
${w^{a}}_{\mu}(x) ={e^{a}}_{\mu}(x)+ {t^{a}}_{\mu}(x)$  defined by 
$\omega^{a}={w^{a}}_{\mu}dx^{\mu}$, where $\omega^{a}$ is the NL SUSY invariant differential forms of 
V-A\cite{VA}, is  invertible and  $s^{\mu \nu}(x) \equiv {w_{a}}^{\mu}(x) w^{{a}{\nu}}(x)$ 
are a unified vierbein and a unified metric tensor in SGM spacetime\cite{KS2}\cite{KS3}. \\
%
%
The SGM action  (\ref{SGM}) is invariant at least under global SO(10), ordinary GL(4R),  
the following new NL SUSY transformation; 
\begin{equation}
\delta \psi^{i}(x) = \zeta^{i} + i \kappa (\bar{\zeta}^{j}{\gamma}^{\rho}\psi^{j}(x)) \partial_{\rho}\psi^{i}(x),
\quad
\delta {e^{a}}_{\mu}(x) = i \kappa (\bar{\zeta}^{j}{\gamma}^{\rho}\psi^{j}(x))\partial_{[\rho} {e^{a}}_{\mu]}(x),
\label{newsusy}
\end{equation} 
where $\zeta^{i}, (i=1,..10)$ is a constant spinor and  $\partial_{[\rho} {e^{a}}_{\mu]}(x) = 
\partial_{\rho}{e^{a}}_{\mu}-\partial_{\mu}{e^{a}}_{\rho}$, \\
the following GL(4R) transformations due to (\ref{newsusy});  
\begin{equation}
\delta_{\zeta} {w^{a}}_{\mu} = \xi^{\nu} \partial_{\nu}{w^{a}}_{\mu} + \partial_{\mu} \xi^{\nu} {w^{a}}_{\nu}, 
\quad
\delta_{\zeta} s_{\mu\nu} = \xi^{\kappa} \partial_{\kappa}s_{\mu\nu} +  
\partial_{\mu} \xi^{\kappa} s_{\kappa\nu} 
+ \partial_{\nu} \xi^{\kappa} s_{\mu\kappa}, 
\label{newgl4r}
\end{equation} 
where $\xi^{\rho}=i \kappa (\bar{\zeta}^{j}{\gamma}^{\rho}\psi^{j}(x))$, 
%
%
%
%
%
%
%
%
%
and the following local Lorentz transformation on $w{^a}_{\mu}$; 
\begin{equation}
\delta_L w{^a}_{\mu}
= \epsilon{^a}_b w{^b}_{\mu}
\label{Lrw}
\end{equation}
with the local  parameter
$\epsilon_{ab} = (1/2) \epsilon_{[ab]}(x)$    
%
%
%
%
%
%
or accordingly on  $\psi$ and $e{^a}_{\mu}$
\begin{equation}
\delta_L \psi(x) = - {i \over 2} \epsilon_{ab}
      \sigma^{ab} \psi,     \quad
\delta_L {e^{a}}_{\mu}(x) = \epsilon{^a}_b e{^b}_{\mu}
      + {\kappa \over 4} \varepsilon^{abcd}
      \bar{\psi} \gamma_5 \gamma_d \psi
      (\partial_{\mu} \epsilon_{bc}).
\label{newlorentz}
\end{equation}
The commutators of two new NL SUSY transformations (\ref{newsusy})  on $\psi(x)$ and  ${e^{a}}_{\mu}(x)$ 
are GL(4R), i.e. new NL SUSY is the square-root of GL(4R), e.g. 
\begin{equation}
[\delta_{\zeta_1}, \delta_{\zeta_2}] \psi
= \Xi^{\mu} \partial_{\mu} \psi,
\quad
[\delta_{\zeta_1}, \delta_{\zeta_2}] e{^a}_{\mu}
= \Xi^{\rho} \partial_{\rho} e{^a}_{\mu}
+ e{^a}_{\rho} \partial_{\mu} \Xi^{\rho},
\label{com1/2-e}
\end{equation}
where 
$\Xi^{\mu} = 2i\kappa (\bar{\zeta}_2 \gamma^{\mu} \zeta_1)
      - \xi_1^{\rho} \xi_2^{\sigma} e{_a}^{\mu}
      (\partial_{[\rho} e{^a}_{\sigma]})$.
They show the closure of the algebra. 
SGM action (\ref{SGM}) is invariant at least under\cite{ST1} 
$[{\rm global\ NL\ SUSY}] \otimes [{\rm local\ GL(4,R)}]$ $\otimes [{\rm local\ Lorentz}] 
\otimes [{\rm global\ SO(N)}], $
which is isomorphic to SO(10)SP corresponding to the linear representation of SGM. 
Now we have written down $N=10$ SUSY theory including graviton, 
which has circumvented the no-go theorem so far. 
\section{Toward Low Energy Theory of SGM }    
The linearlization of such a high nonlinear theory is inevitable to obtain a renormalizable 
field theory which is equivalent.   \\
As a flat space limit of SGM, we have shown that $N=2$ V-A model is equivalent to the spontaneously broken 
$N=2$ linear SUSY(L SUSY)  ${\it vector \ J^{P}=1^{-}}$ gauge supermultiplet model with spontaneously broken SU(2) 
structure\cite{STT2}. 
The linearlization of $N=1$ V-A model has been carried out 
and shows that it is equivalent to 
$N=1$ scalar\cite{IK}\cite{R}\cite{UZ} supermultiplet 
${\it or}$ $N=1$ axial vector\cite{IK}\cite{STT1} 
gauge supermultiplet of L SUSY.   
We conjecture that any global L SUSY (unified) model is equivalent to a NL SUSY model.
These results are favorable to the SGM scenario based upon the composite (eigenstates) nature of all 
elementary particles except graviton.    \\
Now we show explicitly by the heuristic and practical arguments 
that for $N = 2$ SUSY 
a SUSY invariant relation between component fields 
of a vector supermultiplet of L SUSY 
and N-G fermions of the V-A model 
of NL SUSY is written by using only three arbitrary 
dimensionless parameters, which can be recast 
as the vacuum expectation values of auxiliary fields 
in the vector supermultiplet of L SUSY. 
We denote in this paper the component fields of an $N = 2$ U(1) 
gauge supermultiplet \cite{PF} as follows; 
namely, $A$ and $B$ for two physical scalar fields, 
$A_a$ for a U(1) gauge field 
and $\lambda^i \ (i = 1, 2)$ for two Majorana spinors 
in addition to $F$, $G$ and $D$ for three auxiliary scalar fields 
at least for a free vector supermultiplet 
required from the mismatch of the degrees of freedom 
between bosonic and fermionic physical fields. 
The component fields indeed belong to representations 
of a rigid SU(2) \cite{PF}; namely, $\lambda^i$ and ($F$, $G$, $D$) 
belong to representations {\bf 2} and {\bf 3} of SU(2) 
respectively while other fields are singlets. 
In Ref.\cite{STT2} we have linearized $N = 2$ NL SUSY 
in the manifestly SU(2) covariant form.
The L SUSY transformations of these component fields 
generated by constant (Majorana) spinor parameters $\zeta^i$ are  \\
%
${\delta A = \bar\zeta^1 \lambda^1 + \bar\zeta^2 \lambda^2, \
\delta B = i \bar\zeta^1 \gamma_5 \lambda^1 
+ i \bar\zeta^2 \gamma_5 \lambda^2, \
\delta A_a = - i \bar\zeta^1 \gamma_a \lambda^2 
+ i \bar\zeta^2 \gamma_a \lambda^1}$,  \\
${\delta \lambda^1 = \{ (F + i \gamma_5 G) 
- i \!\!\not\!\partial (A + i \gamma_5 B) \} \zeta^1 
- i F_{ab} \sigma^{ab} \zeta^2 + i \gamma_5 \zeta^2 D}$, \\
${\delta \lambda^2 = \{ (F - i \gamma_5 G) 
- i \!\!\not\!\partial (A + i \gamma_5 B) \} \zeta^2 
+ i F_{ab} \sigma^{ab} \zeta^1 + i \gamma_5 \zeta^1 D }$, \\
${\delta F = - i \bar\zeta^1 \!\!\not\!\partial \lambda^1 
- i \bar\zeta^2 \!\!\not\!\partial \lambda^2, \
\delta G = \bar\zeta^1 \gamma_5 \!\!\not\!\partial \lambda^1 
- \bar\zeta^2 \gamma_5 \!\!\not\!\partial \lambda^2, \
\delta D = \bar\zeta^1 \gamma_5 \!\!\not\!\partial \lambda^2 
+ \bar\zeta^2 \gamma_5 \!\!\not\!\partial \lambda^1}$,  \\
%
which satisfy a closed off-shell  algebra.  \\
On the other hand, in the $N = 2$ V-A model 
we have a NL SUSY transformation laws of (Majorana) N-G fermions 
$\psi^i$ generated by $\zeta^i$, 
\be
\delta \psi^i = {1 \over \kappa} \zeta^i 
- i \kappa (\bar\zeta^j \gamma^a \psi^j) \partial_a \psi^i, 
\label{NLSUSY}
\ee
where and hereafter $\kappa$ is a constant whose dimension is $({\rm mass})^{-2}$. 
Eq.(\ref{NLSUSY}) also satisfies the off-shell commutator algebra without 
a U(1) gauge transformation. 

From above L and NL SUSY transformations 
%
%
a SUSY invariant relation between the component fields 
of the $N = 2$ vector supermultiplet and the N-G fermion fields 
$\psi^i$ is obtained at the leading orders of $\kappa$ 
as follows: Indeed, adopting an ansatz 
\begin{equation}
\lambda^1 = (\xi + i \theta \gamma_5) \psi^1 
+ (\eta + i \varphi \gamma_5) \psi^2 + ...\ , \
\lambda^2 = (\xi' + i \theta' \gamma_5) \psi^1 
+ (\eta' + i \varphi' \gamma_5) \psi^2 + ...\ . 
\label{ansatz}
\end{equation}
with $\xi, \eta, \theta, \varphi, \xi', \eta', \theta'$ 
and $\varphi'$ being eight arbitrary real parameters 
which are the most general one for the dimensionless case, 
we substitute (\ref{ansatz}) and (\ref{NLSUSY}) 
into the L SUSY transformations 
and equate them as done in \cite{R}. 
Then we immediately obtain the relation between the bosonic fields 
$A$, $B$, $A_a$, $F$, $G$ and $D$ of the linear supermultiplet 
and the N-G fermions $\psi^i$ at the leading orders of $\kappa$, 
provided that the real parameters in (\ref{ansatz}) 
are restricted to 
\begin{equation}
\eta = 0 = \xi', \
\eta' = \xi, \ \ \ \theta' = \varphi, \
\ \ \ \varphi' = - \theta. 
\label{cnstr}
\end{equation}
The results are 
\begin{eqnarray}
&\!\!\! &\!\!\! 
A = {1 \over 2} \kappa \ \xi \ (\bar\psi^1 \psi^1 + \bar\psi^2 \psi^2) 
+ {i \over 2} \kappa \ \theta 
\ (\bar\psi^1 \gamma_5 \psi^1 - \bar\psi^2 \gamma_5 \psi^2) 
+ i \kappa \ \varphi \ \bar\psi^1 \gamma_5 \psi^2 + ...\ , 
\label{co-A} \\
&\!\!\! &\!\!\! 
B = {i \over 2} \kappa \ \xi \ (\bar\psi^1 \gamma_5 \psi^1 
+ \bar\psi^2 \gamma_5 \psi^2) 
- {1 \over 2} \kappa \ \theta \ (\bar\psi^1 \psi^1 - \bar\psi^2 \psi^2) 
- \kappa \ \varphi \ \bar\psi^1 \psi^2 + ...\ , \\
&\!\!\! &\!\!\! 
A_a = - i \kappa \ \xi \ \bar\psi^1 \gamma_a \psi^2 
+ \kappa \ \theta \ \bar\psi^1 \gamma_5 \gamma_a \psi^2 
- {1 \over 2} \kappa \ \varphi \ (\bar\psi^1 \gamma_5 \gamma_a \psi^1 
- \bar\psi^2 \gamma_5 \gamma_a \psi^2) + ...\ , 
\label{co-Aa} \\
&\!\!\! &\!\!\! 
\lambda^1 = (\xi + i \theta \gamma_5) \psi^1 
+ i \varphi \gamma_5 \psi^2 + ...\ , 
\label{co-l11} \\
&\!\!\! &\!\!\! 
\lambda^2 = (\xi - i \theta \gamma_5) \psi^2 
+ i \varphi \gamma_5 \psi^1 + ...\ , 
\label{co-l21} \\
&\!\!\! &\!\!\! 
F = \xi \ \left\{ {1 \over \kappa} 
- i \kappa (\bar\psi^1 \!\!\not\!\partial \psi^1 
+ \bar\psi^2 \!\!\not\!\partial \psi^2) \right\} 
- \kappa \ \theta \ (\bar\psi^1 \gamma_5 \!\!\not\!\partial \psi^1 
- \bar\psi^2 \gamma_5 \!\!\not\!\partial \psi^2) 
\nonumber \\[.5mm] 
&\!\!\! &\!\!\! \hspace{5mm} 
- \kappa \ \varphi \ \partial_a (\bar\psi^1 \gamma_5 \gamma^a \psi^2) 
+ ...\ , 
\label{co-F} \\
&\!\!\! &\!\!\! 
G = \theta \ \left\{ {1 \over \kappa} 
- i \kappa (\bar\psi^1 \!\!\not\!\partial \psi^1 
+ \bar\psi^2 \!\!\not\!\partial \psi^2) \right\} 
+ \kappa \ \xi \ (\bar\psi^1 \gamma_5 \!\!\not\!\partial \psi^1 
- \bar\psi^2 \gamma_5 \!\!\not\!\partial \psi^2) 
\nonumber \\[.5mm] 
&\!\!\! &\!\!\! \hspace{5mm} 
- i \kappa \ \varphi \ \partial_a (\bar\psi^1 \gamma^a \psi^2) + ...\ , \\
&\!\!\! &\!\!\! 
D = \varphi \ \left\{ {1 \over \kappa} 
- i \kappa (\bar\psi^1 \!\!\not\!\partial \psi^1 
+ \bar\psi^2 \!\!\not\!\partial \psi^2) \right\} 
+ \kappa \ \xi \ \partial_a (\bar\psi^1 \gamma_5 \gamma^a \psi^2) 
\nonumber \\[.5mm]
&\!\!\! &\!\!\! \hspace{5mm} 
+ i \kappa \ \theta \ \partial_a (\bar\psi^1 \gamma^a \psi^2) + ...\ , 
\label{co-D}
\end{eqnarray}
%
in which the three arbitrary real parameters $\xi$, $\theta$ 
and $\varphi$ are involved. The first term 
$- i \kappa \ \xi \ \bar\psi^1 \gamma_a \psi^2$ 
in Eq.(\ref{co-Aa}) shows 
the vector nature of the U(1) gauge field 
as shown in \cite{STT2}. 
Also Eqs.(\ref{co-F}) to (\ref{co-D}) 
for the auxiliary fields $F$, $G$ and $D$ 
have the form which is proportional to a determinant 
$\vert w \vert = {\rm det}(w{^a}_b)$ in the $N = 2$ V-A model 
(with $w{^a}_b$ being defined by 
$w{^a}_b = \delta{^a}_b + t{^a}_b$ and 
$t{^a}_b = - i \kappa^2 \bar\psi^i \gamma^a \partial_b \psi^i$) 
plus total derivative terms at least at the leading 
orders of $\kappa$: namely, 
%
$F = (\xi/\kappa)$ $\times \ [ \ {\rm leading\ terms\ of\ } \vert w \vert \ ] 
+ [ \ {\rm tot.\ der.} \ ] + ... \ ,$ etc. 
%
In addition, the first terms in Eqs.(\ref{co-F}) to (\ref{co-D}) 
or the SUSY transformations of Eqs.(\ref{co-l11}) and (\ref{co-l21}) 
show that $\xi/\kappa$, $\theta/\kappa$ and $\varphi/\kappa$ 
correspond to the vacuum expectation values 
of the auxiliary fields $F$, $G$ and $D$. 

We can continue to obtain higher order terms in the SUSY invariant 
relations: After some calculations we obtain the relation between 
$\lambda^i$ and the N-G fermion fields $\psi^i$ at $O(\kappa^2)$ as 
\begin{eqnarray}
\lambda^1 = 
&\!\!\! &\!\!\!
(\xi + i \theta \gamma_5) \psi^1 + i \varphi \gamma_5 \psi^2  
- \ {i \over 2} \kappa^2 \ \xi 
\ \{ (\bar\psi^1 \!\!\not\!\partial \psi^1) \psi^1 
- (\bar\psi^1 \gamma_5 \!\!\not\!\partial \psi^1) \gamma_5 \psi^1 \nonumber \\[.5mm]
&\!\!\! &\!\!\! + (\bar\psi^1 \partial_a \psi^1) \gamma^a \psi^1 
+ (\bar\psi^1 \gamma_5 \partial_a \psi^1) \gamma_5 \gamma^a \psi^1 \} 
+ ... \ \ , 
\label{co-l12}
\end{eqnarray}
and $\lambda^2$ is obtained by exchanging the indices 1 and 2 
and by replacing $\theta$ with $-\theta$ in Eq.(\ref{co-l12}). 
We can also construct the SUSY invariant relation 
with respect to the bosonic fields of the linear supermultiplet 
at $O(\kappa^3)$ \cite{STT2}. 
In principle we can further continue to obtain higher order terms 
in the SUSY invariant relation following this approach. 
However, it will be more useful to extend  the  superfield formalism 
 Refs.\cite{IK}\cite{UZ}\cite{STT1} to  $N = 2$. 
Remarkably, Eqs.(\ref{co-A}) to (\ref{co-D}) 
(and also (\ref{co-l12}), etc.) 
reduce to that of the $N = 1$ SUSY by imposing, e.g. $\psi^2 = 0$: 
Indeed, when $\xi = 1$ and $\theta = \varphi = 0$, 
they becomes that of the scalar supermultiplet obtained in Ref.\cite{R}. 
When $\varphi = 1$ and $\xi = \theta = 0$, 
they reduce to that of the U(1) gauge supermultiplet obtained 
in Refs.\cite{IK}\cite{STT1}.  \\
Now we consider a  action which is invariant under L SUSY.  
\begin{eqnarray}
S_{\rm lin} = 
&\!\!\! &\!\!\! 
\int d^4 x \left[ {1 \over 2} (\partial_a A)^2 
+ {1 \over 2} (\partial_a B)^2 - {1 \over 4} F^2_{ab} 
+ {i \over 2} \bar\lambda^i \!\!\not\!\partial \lambda^i
+ {1 \over 2} (F^2 + G^2 + D^2) \right. \nonumber \\[.5mm]
&\!\!\! &\!\!\! 
\hspace{1.2cm} \left. 
- {1 \over \kappa} (\xi F + \theta G + \varphi D) \right], 
\label{Lact}
\end{eqnarray}
where $\xi$, $\theta$ and $\varphi$ are three arbitrary 
real parameters satisfying $\xi^2 + \theta^2 + \varphi^2 = 1$. 
The last three terms proportional to $\kappa^{-1}$ 
is an analogue of the Fayet-Iliopoulos $D$ term in the $N = 1$ 
theories \cite{FI}. 
The field equations for the auxiliary fields $F$, $G$ or $D$ 
are $F = \xi/\kappa$, $G = \theta/\kappa$ 
or $D = \varphi/\kappa$ indicating a spontaneous SUSY breaking. 
Substituting (\ref{co-A}) to (\ref{co-D}) 
into the linear action $S_{\rm lin}$ of (\ref{Lact}), 
we can show immediately that $S_{\rm lin}$ coincides with 
the following V-A action $S_{\rm VA}$  up to and including $O(\kappa^0)$; 
namely, 
$S_{\rm VA} = 
- {1 \over {2 \kappa^2}} 
\int d^4 x \ \vert w \vert = 
- {1 \over {2 \kappa^2}} \int d^4 x 
[ 1 + t{^a}_a + ... \ ], $
%
which is invariant under (\ref{NLSUSY}). \\
We note that the linearization of $N = 2$ SUSY in this paper 
can be discussed as a manifestly (rigid) SU(2) invariant form 
\cite{STT2}, which gives more concise expressions of the SUSY 
invariant relation (\ref{co-A}) to (\ref{co-D}) 
(and also (\ref{co-l12}), etc.). 
In these arguments, adopting the general ansatz (\ref{ansatz}) 
having the eight real dimensionless parameters with $\kappa^0$, 
we have explicitly shown that for $N = 2$ SUSY the SUSY invariant relation 
(\ref{co-A}) to (\ref{co-D}) (and also (\ref{co-l12}), etc.) 
is written by using only three arbitrary parameters, 
which can be recast as the vacuum expectation values 
of the auxiliary fields in the vector supermultiplet. 
These heuristic arguments are practical and show more general assumptions 
adopted for obtaining the SUSY invariant relation. \\
The analysis by using the NL SUSY superfield in curved spacetime\cite{WB} 
may be useful to carry out the computations to all orders and make the arguments 
transparent. \\
From those arguments on the linearization of $N = 1$ and $N = 2$ SUSY, 
we speculate that any renormalizable ($N$-exteded) global L SUSY 
(interacting) model is equivalent to the ($N$-extended) V-A model 
despite the difference of the number of the dynamical degrees of freedom. 
These results support the SGM scenario \cite{KS1,KS2} 
which is a composite model of matter based on the global NL SUSY (generalized V-A\cite{W}) model in curved spacetime. \\
It is interesting that the appearance of vector (not axial) gauge field as a composite 
necessitates $N = 2$ NL SUSY, i.e. SU(2) structure (and its spontaneous breakdown to U(1)) in L SUSY. 
The compositeness of all elementary particles, i.e. composite picture of SGM  may explain  
$SU(2) (\times U(1))$ gauge structure in SM.  \\
As for the linearlization of SGM, we have recently obtained the SUSY invariant relations 
between ${e^{a}}_{\mu}(x)$, $\psi_{\mu}(x)$ and  N-G field $\psi(x)$, e.g. 
\be
\psi_{\mu}(x)={\sqrt \kappa \over 4}\gamma_{a}\gamma^{\rho}\psi(x)\partial_{[\rho} {e^{a}}_{\mu]}(x), 
\ee
which produce the closed algebra \cite{STS1}. 
The details of SGM case will appear soon. 
The cosmology of SGM and SGM with extra dimensions, ...,etc. are open. 
\bigskip

{\small The work of M. Sawaguchi is supported in part by the special research project of 
High-Tech Research Center of SIT}.

\end{document}